\documentstyle[12pt]{article}
\begin{document}
\baselineskip=.3in
\newcommand{\be}{\begin{equation}}
\newcommand{\ee}{\end{equation}}
\newcommand{\bea}{\begin{eqnarray}}
\newcommand{\eea}{\end{eqnarray}}
\newcommand{\lagrangian}{{\cal L}}
\newcommand{\fumn}{F^{\mu\nu}}
\newcommand{\flmn}{F_{\mu\nu}}
\newcommand{\eumnr}{\varepsilon^{\mu\nu\rho}}
\newcommand{\dln}{\partial_{\nu}}
\newcommand{\dlm}{\partial_{\mu}}
\newcommand{\dli}{\partial_i}
\newcommand{\dlj}{\partial_j}
\newcommand{\dlt}{\partial_t}
\newcommand{\dlz}{\partial_0}
\newcommand{\dlo}{\partial_1}
\newcommand{\inttd}{\int\! d^2x}
\newcommand{\intr}{\int_0^\infty \! dr}
\newcommand{\lrarrow}{\longrightarrow}
\renewcommand{\thesection}{\Roman{section}.}
\title{\LARGE {\bf Self-dual Vortices in the Generalized
\protect\\[1mm] Abelian Higgs Model with Independent Chern-Simons Interaction}}
\author{Chanju Kim \\
{\normalsize\it Department of Physics, Seoul National University}\\
{\normalsize\it Seoul 151-742, Korea}}
\date{}
\maketitle

\def\thepage{\protect\raisebox{0ex}{\ }SNUTP 92--79
      \protect\hspace{-28mm}\protect\raisebox{-2.4ex}{hepth@xxx/9209110}}
\thispagestyle{headings}
\markright{\thepage}
\begin{center}
{\bf Abstract}\\[2mm]
\end{center}
\indent\indent
Self-dual vortex solutions are studied in detail in the generalized abelian
Higgs model with independent Chern-Simons interaction. For special choices
of couplings, it reduces to a Maxwell-Higgs model with two scalar fields,
a Chern-Simons-Higgs model with two scalar fields, or other new models.
We investigate the properties of the static solutions and perform detailed
numerical analyses. For the
Chern-Simons-Higgs model with two scalar fields in an asymmetric phase, we
prove
the existence of multisoliton solutions which can be viewed as
hybrids of Chern-Simons vortices and $CP^1$ lumps. We also discuss solutions
in a symmetric phase with the help of the corresponding exact solutions
in its nonrelativistic
limit. The model interpolating all three models---Maxwell-Higgs,
Chern-Simons-Higgs, and $CP^1$ models--- is discussed briefly.
Finally we study the possibility of vortex solutions with half-integer
vorticity in the special case of the model. Numerical results are negative.
\newpage

\pagenumbering{arabic}
\thispagestyle{plain}
\section{INTRODUCTION}
\ \indent
In the last few years there has been much interest in the
(2+1)-dimensional abelian Higgs models and the $CP^1$ model with a
Chern-Simons term in the gauge field action, partly because of their
possible relevance in condensed matter physics
\cite{polyakov}. These models allow classical
vortexlike solutions and the analysis of them has become quite active recently
\cite{khare,cp1}. Here particularly interesting are the so-called self-dual
systems.
Since the discovery of the self-dual charged vortex solutions in relativistic
pure Chern-Simons-Higgs system \cite{hkp}, several other self-dual
systems including the Chern-Simons term are now known \cite{jp,llm,lmr}.
Of some interest among these is the
generalized self-dual abelian Higgs model with independent Chern-Simons
interaction, discussed in Refs. \cite{llm,lmr}. It has two abelian gauge
fields,
one of which has only the Maxwell term as its kinetic term and the other only
the Chern-Simons term. It may be considered as a realistic model
if one identifies
the Maxwell field as the electromagnetic field and the Chern-Simons field as an
effective field arising in a condensed matter system.
Besides, as we will see below,
it contains a linearized version of the $CP^1$ model with or
without a Chern-Simons term as special cases and therefore
interpolates abelian Higgs systems and the $CP^1$ model.
The present paper is devoted to the detailed study of static solutions in
this model, which was not given in Ref. \cite{llm}.

This model consists of two complex scalar fields, a neutral scalar field, and
two gauge fields. Each complex scalar field couples to both gauge fields and
so there are four gauge couplings in general. But, with some special
choices made for these couplings, the model reduces to simpler systems.
For example, for some particular choices \cite{llm} it becomes a decoupled sum
of self-dual Maxwell Higgs and self-dual Chern-Simons Higgs systems.
More interestingly, it is possible to obtain a system in which
two scalar fields interact with a Maxwell or Chern-Simons gauge field
(but not with both),
while possessing an additional global $SU(2)$ symmetry.
This kind of model with a Maxwell gauge field has attracted
much attention recently \cite{gibbons} because it admits topological vortex
solutions in spite of having a simply connected vacuum manifold, and
a class of solutions obtained here
interpolates ordinary Ginzburg-Landau vortices and $CP^1$ lumps \cite{gibbons}.
We consider the corresponding model with a Chern-Simons gauge field, and find
that it has nontopological vortex solutions in a
symmetric phase as well as topological vortices in an asymmetric phase (as
in the Maxwell case).
Again a class of solutions obtained in this case approximate $CP^1$ lumps,
and the existence of
multivortex solutions can be shown by adapting the method of Wang \cite{wang}.
There are also more general self-dual systems based on the same set of
fields but with no additional global symmetry.

If both gauge fields couple to scalar fields nontrivially, a new kind of models
are obtained \cite{llm}. If one of the two complex scalar fields
decouples and the other
couples to both gauge fields, only a broken phase is possible and topological
vortices exist. We find that the nature of the vortex solutions is very similar
to those of the Ginzburg-Landau system, the Chern-Simons field playing a minor
role only.
We will briefly discuss its extension to the two complex-scalar-field
case in such a way that there is an additional global $SU(2)$ symmetry.
Complicated but interesting is the case (f) of Ref. \cite{llm}, where all
fields couple to one another nontrivially. In ref \cite{llm},
it was noted that there might be solutions with
a half-integer vorticity (as regards each of the two gauge fields).
We will report the result of our numerical work, which is negative.

This paper is organized as follows. In Sec. II, we briefly review the model
to be considered by us, following Ref. \cite{llm}.
Section III is the major part of the paper. We study
static solutions for various
interesting cases of this model, based on analytical and numerical means.
We summarize our findings in Sec. IV.
In the appendix, a $U(1)$ self-dual Maxwell-Chern-Simons Higgs
system with additional global $SU(2)$ symmetry is given.
This model might be interesting because the solutions
interpolate all three cases---Ginzburg-Landau, Chern-Simons, and $CP^1$
solitons.

\section{THE MODEL}
\ \indent We consider the (2+1)-dimensional system described by the
following Lagrangian \cite{llm}
\be \label{1}
\lagrangian=-\frac{1}{4}\fumn\flmn+\frac{1}{2} \kappa\eumnr a_\mu \dln a_\rho
            +|D_\mu\chi|^2+|D_\mu\psi|^2+\frac{1}{2}(\dlm N)^2 -
U(\chi,\psi,N),
\ee
where $D_\mu\chi=[\dlm-i(e_1a_\mu+e_2A_\mu)]\chi,$
$ D_\mu\psi=[\dlm-i(e_3a_\mu+e_4A_\mu)]\psi$,
and
\bea                  \label{2}
U(\chi, \psi, N)&=&\frac{1}{2}(e_2|\chi|^2+\sigma_1 e_4|\psi|^2-u^2)^2
   +\frac{e_1^2}{\kappa^2} (e_1|\chi|^2+\sigma_1 e_3|\psi|^2
   -\sigma_2 \frac{\kappa e_2}{e_1} N-v^2)^2\nonumber\\
&&+\frac{e_3^2}{\kappa^2}|\psi|^2(e_1|\chi|^2+\sigma_1 e_3|\chi|^2
-\sigma_2 \frac{\kappa e_4}{e_3} N-v^2)^2.
\eea
Here, $\sigma_1$ and $\sigma_2$ may assume the values $\pm 1$ independently,
$\chi$ and $\psi$ denote complex scalar fields with couplings to both Maxwell
and Chern-Simons gauge fields (i.e., $A_\mu$ and $a_\mu$), while $N$
is a neutral scalar.
The specific form of the potential (\ref{2}) leads to a self-dual system
for general $\sigma_1, \sigma_2=\pm 1$, and only the case of
$\sigma_1=\sigma_2=+1$ was considered in Ref. \cite{llm}.
We shall now briefly summarize the results of Ref. \cite{llm} and extend them
to  generally allowed values of $\sigma_1, \sigma_2$.
The variation of $a_0$ and $A_0$ leads to two Gauss laws
\bea
\kappa f_{12}&=&e_1J^0_\chi+e_3J^0_\psi,   \label{3}\\
\dli F^{i0}&=&e_2J^0_\chi+e_4J^0_\psi,     \label{4}
\eea
where $f_{\mu\nu}=\dlm a_\nu-\dln a_\mu$,
$J^\mu_\chi=-i(\chi^*D^\mu\chi-D^\mu\chi^*\chi)$, and
$J^\mu_\psi=-i(\psi^*D^\mu\psi-D^\mu\psi^*\psi)$. For finite energy
configuration, Eqs. (\ref{3}) and (\ref{4}) imply relations among the
magnetic flux $\Phi_a=\inttd f_{12}$ and charges $Q_\chi=\inttd J^0_\chi$
and $Q_\psi=\inttd J^0_\psi$, viz.,
\bea
\kappa\Phi_a&=&e_1Q_\chi+e_3Q_\psi, \label{5}\\
0&=&e_2Q_\chi+e_4Q_\psi. \label{6}
\eea
There is no restriction on $\Phi_A=\inttd F_{12}$. The energy functional is
\bea   \label{7}
E&=&\inttd\left[\frac{1}{2}F^2_{i0}+\frac{1}{2}F^2_{12}+|D_0\chi|^2+
|D_0\psi|^2\right.\nonumber\\
&&\left.+|D_i\chi|^2+|D_i\psi|^2
+\frac{1}{2}(\dlz N)^2+\frac{1}{2} (\dli N)^2+U\right].
\eea
After employing the Gauss laws and integrating by parts, one has the bound
\be
E\geq|v^2\Phi_a+u^2\Phi_A|,
\ee
where the equality holds if and only if the following self-duality
equations are satisfied:
\bea
&&(D_1\pm iD_2)\chi = (D_1\pm i\sigma_1 D_2)\psi=0,\nonumber\\
&&F_{12}\pm(e_2|\chi|^2+\sigma_1 e_4|\psi|^2-u^2)=0,\nonumber\\
&&F_{i0}\pm\sigma_2 \dli N=0 \label{9},\\
&&D_0\chi\pm i\frac{e_1}{\kappa}\chi(e_1|\chi|^2+\sigma_1 e_3|\psi|^2
-\sigma_2 \frac{\kappa e_2}{e_1} N-v^2)=0, \nonumber\\
&&D_0\psi\pm i\frac{e_3}{\kappa}\psi(e_1|\chi|^2+\sigma_1 e_3|\psi|^2
-\sigma_2 \frac{\kappa e_4}{e_3} N-v^2)=0.\nonumber
\eea
Note that if $\sigma_1=-1$, the first equation shows that one of the
two complex scalar fields is self-dual while the other is antiself-dual.
Also, depending on the value of $\sigma_2$,  the sign in front of $N$
gets changed.  However, each choice of $(\sigma_1, \sigma_2)$
corresponds to a different potential as
seen in Eq. (\ref{2}) and all signs in Eq. (\ref{9}) are correlated in a
given theory. Each self-dual system for given $\sigma_1$ and $\sigma_2$ is in
fact not much different from the others. From Eqs. (\ref{1}), (\ref{2}) and
(\ref{9}), we see that the system with $\sigma_1=-1$ is transformed into
that with $\sigma_1=+1$ under $e_3\rightarrow -e_3$, $e_4\rightarrow -e_4$
and $\psi\rightarrow\psi^*$, and into the system with $\sigma_2=-1$
by $N\rightarrow -N$.
Therefore, we put $\sigma_1=\sigma_2=+1$ from now on, except at the end of
Sec. III{\bf A} where we will give some further comments.

We are interested in static solitonlike solutions of Eq. (\ref{9}). Since
there are four different gauge couplings, the model has a rich variety and the
nature of solutions depends on the choice of these couplings.
It should be thus desirable to consider separately various
cases corresponding to
special choices of couplings. For certain choices of couplings some of
known self-dual systems, including the model of some recent interest
\cite{gibbons}, are recovered. For other choices we encounter models which
have not been studied in detail so far.

\section{SPECIAL CASES}
\subsection*{A. {\boldmath $e_1=e_3=0$}}
\ \indent If the couplings associated with $a_\mu$, i.e., $e_1$ and $e_3$,
are identically zero, then $a_\mu$ decouples and we have a self-dual
Maxwell-Higgs system with two complex scalar fields and a neutral scalar.
This is a generalization of the well-known self-dual Ginzburg-Landau
model \cite{landau}. Furthermore, from Eq. (\ref{9}), the equation for $N$
becomes $\nabla^2N=2(e_2^2|\chi|^2+e_4^2|\psi|^2)N$, which means that $N$
may be taken to be zero identically.
Here the potential is effectively
\be \label{10}
U=\frac{1}{2}(e_2|\chi|^2+e_4|\psi|^2-u^2)^2.
\ee
If one further restricts oneself to the case $e_2=e_4$, this model will
possess additional global $SU(2)$ symmetry and become in fact identical
to the one recently considered by Vachaspati
{\it et al.}, Hindmarsh, and Gibbons {\it et al.} \cite{gibbons}. For
general $e_2$ and $e_4$, there is no additional global symmetry and we
study this general case below.

{}From Eqs. (\ref{7})and (\ref{10}), finiteness of energy requires that

\be \label{11}
r\rightarrow\infty :\hspace{10mm}e_2|\chi|^2+e_4|\psi|^2\lrarrow u^2
\ee
as well as
\be \label{12}
r\rightarrow\infty :\hspace{10mm} D_i\chi \lrarrow 0,\hspace{5mm}
D_i\psi \lrarrow 0.
\ee
Eq. (\ref{11}) tell us that
on the circle at infinity, scalar fields must lie on an
ellipsoid\footnote{For the moment, we consider the case $e_2>0,$ $e_4>0$
only.} in the 4-dimensional internal space; then, Eq. (\ref{12}) further
restricts the scalar fields to be at
most a pure phase there. Hence even if the vacuum manifold is simply connected,
there can be vortex solutions characterized by a winding number
\cite{gibbons}\footnote{In fact this phenomenon
has been known since seventies even if such model has not been explicitly
constructed \cite{coleman}.}.

Now we consider the self-duality equations (\ref{9}). On points where $\chi$,
$\psi$ are nonzero, the first equation may be rewritten as
$$
i(A_1+iA_2)=\frac{1}{e_2}(\partial_1+i\partial_2)\ln\chi
           =\frac{1}{e_4}(\partial_1+i\partial_2)\ln\psi,
$$
where we chose the upper sign. Then
$$
(\dlo+i\partial_2)\ln\frac{\psi^{e_2/e_4}}{\chi}=0,
$$
so that
\be
w(z)\equiv\frac{\psi^{e_2/e_4}}{\chi}  \label{13}
\ee
is locally analytic in $z=x+iy$. Denoting $z_r(r=1,\ldots,n_1)$ and $ z'_r
(r=1,\ldots,n_2)$ to be the zeros of $\chi$ and $\psi$ respectively, we
may then write \cite{gibbons,taubs}
\be
w(z)=\frac{\prod_{r=1}^{n_2}(z-z'_r)^{e_2/e_4}}{
     \prod_{r=1}^{n_1}(z-z_r)}h(z),      \label{14}
\ee
where $h(z)$ has no zero or pole and is regular everywhere. Since $\chi$ and
$\psi$ should approach vacuum values as $r\rightarrow\infty$, $h(z)$ is at
most a constant, say $q$,
by Liouville's theorem in complex variable theory. Note that $w(z)$ is
multivalued in general, but it is not problematic as long as $\chi$ and
$\psi$ are single-valued functions.
Now let
\be
r\lrarrow\infty:\hspace{10mm}\left(\begin{array}{c}\chi\\
\psi\end{array}\right)
\lrarrow\left(\begin{array}{c}\chi_0\\ \psi_0\end{array}\right),
\ee
where $\chi_0$ and $\psi_0$ are some fixed constants satisfying
$e_2|\chi_0|^2+e_4|\psi_0|^2=u^2$.
If $e_2=e_4$, all vacuums are equivalent because of global $SU(2)$ symmetry
and hence it suffices to choose any convenient values for $\chi_0$ and
$\psi_0$,
for example, $\psi_0=0$ and $|\chi_0|=u/\sqrt{e_2}$. However, if $e_2\neq e_4$,
various choices for the vacuum values are inequivalent and
one must consider in particular the case with $\chi_0\psi_0\neq 0$.
In the latter case, from Eq. (\ref{13}), we have
$w(z)\rightarrow$ finite as $r\rightarrow\infty$, which implies in view of
Eq. (\ref{14}) that $e_2/e_4=n_1/n_2$. In other words, regular vortex
solutions may exist only if $e_2/e_4$ is rational and the ratio of the number
of zeros for $\chi$ and $\psi$, $n_1/n_2$, must have the same fixed value
equal to $e_2/e_4$.  The flux is then given by
\be
\Phi\equiv \oint_{r=\infty}dx^iA_i=\frac{2\pi n_1}{e_2}=\frac{2\pi n_2}{e_4}.
\ee
On the other hand, if one of the vacuum values, say $\psi_0$, is zero we
can have a totally different behaviors for the $e_2\neq e_4$ case, which is
absent for the case $e_2=e_4$. From Eq. (\ref{13}), $w\rightarrow 0$ as
$r\rightarrow \infty$ and this implies
only the condition $e_2n_2<e_4n_1$. There is no need for $e_2/e_4$ to be
rational here and, except the above inequality, no condition on $n_1/n_2$,
either. This can be understood once if one notices that the vacuum manifold for
$\psi$ is trivial in this case. Now suppose that
$$
\psi\lrarrow\frac{1}{r^\alpha}\hspace{7mm}\mbox{as }r\lrarrow\infty.
$$
Then the flux will be, from Eq. (\ref{9}),
\be
\Phi=\frac{2\pi n_1}{e_2}=\frac{2\pi(n_2+\alpha)}{e_4},
\ee
or $\alpha$ is given by
\be
\alpha=\frac{e_4}{e_2}n_1-n_2>0.
\ee

There is a further non-trivial equation to be satisfied by $|\chi|$ ( or by
$|\psi|$). Eliminating gauge fields from the self-duality equations (\ref{9}),
we have in fact
\be \label{sf1}
\nabla^2\ln|\chi|^2-2e_2(e_2|\chi|^2+e_4|\psi|^2-u^2)=4\pi\sum_{r=1}^{n_1}
\delta({\bf x}-{\bf x}_r),
\ee
where $|\psi|=(|w||\chi|)^{e_4/e_2}$.
If $e_2=e_4\equiv e$, we can simplify this equation as \cite{gibbons}
\be \label{sf2}
\nabla^2\eta+2eu^2(1-e^\eta)=\nabla^2\ln\left(\prod_{r=1}^{n_1}|z-z_r|^2
          +|q|^2\prod_{r=1}^{n_2}|z-z'_r|^2\right),
\ee
where $\eta=\ln [e(|\chi|^2+|\psi|^2)/u^2]$.
Analytic solutions to Eq. (\ref{sf1}) or (\ref{sf2}) are not available
and we should resort to numerical analysis to go
further. The simplest case of $e_2=e_4$, with additional global $SU(2)$
symmetry, gives rise to some interesting consequences, but they were
extensively discussed already in Ref. \cite{gibbons}\footnote{However,
see the next section.}.

For generic case, some numerical solutions have been studied by us for
specific couplings and
vorticities. In Fig. 1(a) we have plotted rotationally symmetric solutions
with unit vorticity when the ratio $p\equiv \frac{e_4}{e_2}$ is equal to 1,
or 2, while
choosing the point $|\chi_0|^2\equiv u^2/2e_2$ on the ellipse of possible
vacuum values. In addition, we studied solutions for other $n$ and $p$ values
and we found that the shapes are more or less the same for different
$p$-values except that the vortex sizes become a little
smaller for larger $p$. In Fig 1(b) we have plotted $e^{\eta/2}$ for the
case $n_1=1$ and $n_2=0$ with $e_2=e_4\equiv e$. Note that the approximation
$e^{\eta/2}=1$ is excellent for $|q|=10$ and good even for $|q|=5$.
Therefore the solutions well approximate
$CP^1$ lumps for large $|q|$, as noted in Ref. \cite{gibbons}\footnote{Readers
may consult discussions given in the case {\bf B} for the relevance of the
$CP^1$ model here.}.
For the number of free parameters in the general solution (which need not
be rotationally symmetric), one can use the index theorem
and this was already done in Ref. \cite{lmr} for the case with
$\chi_0\psi_0\neq 0$. The number is equal to $2(n_1+n_2)$.

Closing the discussions, we comment on the case with $e_2e_4<0$.
{}From Eqs.(\ref{13}) and (\ref{14}), we now find that
$$
\frac{1}{w}=\chi\psi^{|e_2/e_4|}
  =\prod_{r=1}^{n_1}(z-z_r)\prod_{r=1}^{n_2}(z-z'_r)^{|e_2/e_4|}h_0,
$$
where $h_0$ is a constant. Then $n_1$ and $n_2$ should be zero because $1/w$
approaches a finite value as $r\rightarrow\infty$. This implies that the energy
$E=u^2|\Phi_A|$ is zero and there is no nontrivial solution satisfying
the self-duality equations. However, as we mentioned in Sec. II,
there is another self-dual system
which corresponds to $\sigma_1=-1$ in Eq. (\ref{2}), and this
system is essentially identical to that with $e_2e_4>0$ and $\sigma_1=+1$.

\subsection*{B. {\boldmath $e_2=e_4=0$}}
\ \indent
With $e_2$ and $e_4$ set to zero, the Maxwell $U(1)$ field decouples from
matters fields, and so does the neutral field $N$.
It is evidently the Chern-Simons counterpart of the model considered in the
case {\bf A}, and similar analysis can be used to find soliton solutions
allowed for this case. Here we have the potential
\be
U=\frac{1}{2}(e_1^2|\chi|^2+e_3^2|\psi|^2)(e_1|\chi|^2+e_3|\psi|^2-v^2)^2,
\ee
and, if $e_1=e_3$, there is an additional global $SU(2)$ symmetry in
the system. From the self-duality equations (\ref{9}),
we may conclude, just as in the case {\bf A} (see Eq. (\ref{14})), that
\be
w(z)\equiv\frac{\psi^{e_1/e_3}}{\chi}=
       q\frac{\prod_{r=1}^{n_2}(z-z'_r)^{e_1/e_3}}
       {\prod_{r=1}^{n_1}(z-z_r)},  \label{19}
\ee
where $q$ is an arbitrary complex constant and $z_r(r=1,\ldots,n_1)$
and $z'_r(r=1,\ldots,n_2)$ are zeros of $\chi$ and $ \psi$, respectively.
In the present case, however, the situation is more complicated because
there is also a symmetric phase as well as an asymmetric or broken phase;
in the symmetric phase, we have an additional possibility of nontopological
soliton solutions.

First, we concentrate on the topological soliton solutions allowed in the
broken phase, for which the analysis is almost an exact parallel to that of the
case {\bf A}. For example, the statements below Eq. (\ref{14})
apply also to this case with $e_2$ and $e_4$ replaced by $e_1$ and $e_3$,
respectively. In particular, with $e_1=e_3\equiv e$, additional global $SU(2)$
symmetry allows us to assume, without loss of generality, that
\be
r\lrarrow\infty:\hspace{7mm}
\left(\begin{array}{c}\chi\\ \psi\end{array}\right)\lrarrow
    \left(\begin{array}{c}\chi_0\\ 0\end{array}\right),\hspace{5mm}
\left(|\chi_0|^2=\frac{v^2}{e_1}\right),
\ee
and we can follow Gibbons {\it et al.} \cite{gibbons} closely. An immediate
observation is that the function $w(z)\equiv\psi/\chi$ should vanish at
infinity and Eq. (\ref{19}) reduces to
\be
w(z)=\frac{Q_n(z)}{P_n(z)},
\ee
where $P_n(z)=\prod_{r=1}^n(z-z_r),$ and $Q_n(z)$ is
a polynomial of $z$ of order not larger than $n-1$. Further, eliminating
gauge field in Eq. (\ref{9}) yields the following equation for $|\chi|$ (note
that $|\psi|^2=|w|^2|\chi|^2$):
\be  \label{22}
\nabla^2\ln|\chi|^2-\frac{4e^4}{\kappa^2}(|\chi|^2+|\psi|^2)\left(|\chi|^2+
       |\psi|^2-\frac{v^2}{e}\right)=4\pi\sum_{r=1}^{n}\delta({\bf x}-
      {\bf x}_r)
\ee
Choosing to work with dimensionless quantities, we make the replacements
$\chi\rightarrow v\chi/\sqrt{e},$
$\psi\rightarrow v\psi/\sqrt{e},$
$a_\mu\rightarrow 2v^2a_\mu/\kappa$,
and then Eq. (\ref{22}) reads
\be \label{23}
\nabla^2\ln|\chi|^2 -(|\chi|^2 +|\psi|^2)[(|\chi|^2+|\psi|^2)-1]
       =4\pi\sum_{r=1}^{n}\delta({\bf x}-{\bf x}_r).
\ee

To facilitate the analysis of Eq. (\ref{23}), we introduce the quantity
\be  \label{24}
\eta=\ln(|\chi|^2+|\psi|^2)
\ee
so that we can recast Eq. (\ref{23}) as
\be  \label{25}
\nabla^2\eta-e^\eta(e^\eta-1)=\rho,
\ee
where we defined
\be  \label{26}
\rho=\nabla^2\ln(|P_n|^2+|Q_n|^2).
\ee
[Note that $\nabla^2\ln|z-z_r|^2=4\pi\delta({\bf x}-{\bf x}_r)$.]
Since $|\chi|^2+|\psi|^2$ approaches 1 as $r\rightarrow\infty$,
we here expect that $\eta\rightarrow 0$ as $r\rightarrow\infty$.
Following ref. \cite{gibbons}, we define
\be \label{27}
u_1=\nabla^2\ln(|P_n|^2+|Q_n|^2)-\sum_{r=1}^n\ln(|z-z_r|^2+\mu),
\ee
where $\mu(>0)$ is for the moment an arbitrary constant. Setting
$\eta=u_1+{\tilde\eta}$ then gives
\be  \label{28}
\nabla^2{\tilde\eta}-h+e^{u_1}e^{\tilde\eta}(1-e^{u_1}e^{\tilde\eta})=0
\ee
with $h=4\sum_{r=1}^n\mu(|z-z_r|^2+\mu)^{-2}.$ Eq. (\ref{28}) corresponds
to the variational problem associated with the functional
\be \label{29}
{\cal A}({\tilde\eta})=\inttd
 \left[\frac{1}{2}|\nabla {\tilde\eta}|^2+(e^{u_1+{\tilde\eta}}-1)^2
  +2h{\tilde\eta}\right],
\ee
where ${\tilde\eta}$ should vanish sufficiently fast as $r\rightarrow\infty$.
With $Q_n\equiv 0$, this is in fact the action used by Wang \cite{wang} to
prove the existence of multivortex solutions in the self-dual pure Chern-Simons
model with one complex scalar field. His method can be adapted to the present
case. Wang's proof for $Q_n=0$ rests essentially on showing that
${\cal A}({\tilde\eta})$ is a coercive functional and hence
${\cal A}({\tilde\eta})$
has a unique minimum. Other parts differing from the $Q_n=0$ case having
been checked already by Gibbons {\it et al.} \cite{gibbons} for sufficiently
large $\mu$, the only crucial step needed in establishing coercivity is to
estimate the quantity $\inttd e^{2u_1}(e^{\tilde\eta}-1)^2$.
Wang proved for $Q_n=0$ that
\be  \label{30}
\inttd e^{2u_1}(e^{\tilde\eta}-1)^2\geq
 \frac{1}{2}\int\left(\frac{|{\tilde\eta}|}{1+|{\tilde\eta}|}\right)^2-C,
\ee
where  $C$ is a positive constant. However, it is not difficult to see that
the same argument can be used also for the case $Q_n\neq 0$
and hence Eq. (\ref{30}) holds even if $Q_n\neq 0$. We conclude that
for every choice of polynomial $P_n$ and $Q_n$, a unique solution exists and
the scalar fields may be reconstructed from $\eta$ via
\bea  \label{31}
\left(\begin{array}{c}\chi\\ \psi\end{array}\right)&=&
     \frac{1}{\sqrt{1+|w|^2}}\left(\begin{array}{c}1\\ w\end{array}\right)
     e^{\eta/2}\prod_{r=1}^n\frac{z-z_r}{|z-z_r|}\nonumber\\
&=&\frac{1}{\sqrt{|P_n|^2+|Q_n|^2}}
       \left(\begin{array}{c}P_n\\ Q_n\end{array}\right)e^{\eta/2}
\eea
up to gauge transformations. The gauge field $a_i$ is then readily obtained
using Eq. (\ref{9}). The moduli space of solutions is just $C^{2n}$,
the $4n$-dimensional space parametrized by the complex coefficients specifying
the polynomials $P_n$ and $Q_n$\footnote{Explicitly, the free parameters can
be identified with $p_k$, $q_k$ ($k=0, 1, \ldots, n-1)$ when one writes
$P_n(z)=z^n+p_{n-1}z^{n-1}+\cdots +p_1z+p_0 $ and
$Q_n(z)=q_{n-1}z^{n-1}+\cdots + q_1+q_0.$}.
We note that the equation of the form (\ref{31}) has been given for the case
{\bf A} in Ref. \cite{gibbons}. The index theorem analysis of Lee, Min and
Rim \cite{lmr} on the number of zero modes is also consistent with what we
have shown rigorously here.

A characteristic feature of Chern-Simons vortices is that, unlike the
Ginzburg-Landau vortices, they carry nonzero angular momentum.
The angular momentum is
\bea  \label{32}
J&=&-\frac{\kappa}{2e^2}\inttd \varepsilon^{ij}x^i(D_0\chi^*D_j\chi
    +D_0\psi^*D_0\psi+\mbox{c.c.})\nonumber\\
 &=&\!\!\!\inttd\varepsilon^{ij}x^i\frac{f_{12}}{|\chi|^2+|\psi|^2}
    \left[|\chi|^2(a_j-\dlj \arg\chi)+|\psi|^2(a_j-\dlj \arg\psi)\right],
\eea
where we used the self-duality equations on the second expression. Now
suppose we restrict ourselves to solutions with $P_n(z)= z^{n_1}$,
$Q_n(z)=qz^{n_2},$ $(n_2<n_1)$,
where $q$ is a constant, in Eq. (\ref{31}). This corresponds to working with
the ansatz
\bea \label{34}
\chi&=&|\chi|e^{in_1\theta},\nonumber\\
\psi&=&|\psi|e^{in_2\theta},\\
a_i&=&\varepsilon^{ij}\frac{x^j}{r^2}\left(g-\frac{n_1+n_2}{2}\right),\nonumber
\eea
where $|\chi|,$ $|\psi|$ and $g$ are functions of $r$ and $|\psi|/|\chi|=
qr^{n_2-n_1}$. Then Eq. (\ref{32}) becomes
\be  \label{35}
J=\frac{\pi\kappa}{e^2}[g^2(\infty)-g^2(0)]+\frac{\pi\kappa}{e^2}(n_1-n_2)
   \intr g'\frac{r^{2n_1}-|q|^2r^{2n_2}}{r^{2n_1}+|q|^2r^{2n_2}}.
\ee
Obviously, $g(0)=(n_1+n_2)/2 $ from Eq. (\ref{34}), while the finite
energy condition requires that $g(\infty)=-\frac{1}{2}(n_1-n_2)$. Therefore,
the first term on the right hand side of Eq. (\ref{35}) is equal to
$-\pi\kappa n_1n_2/e^2$.
But the second term cannot be
explicitly evaluated unless $\psi$ vanishes identically. For the special cases
corresponding to $\psi\equiv 0$, we have
\be \label{38}
J=-\frac{\pi\kappa n_1^2}{e^2},
\ee
which is in agreement with the result of Ref. \cite{hkp}.

Let us consider the case with $n_1=1$ and $n_2=0$ in more detail. For this
case we may write
\be \label{39}
\left(\begin{array}{c}\chi\\ \psi\end{array}\right)
=\frac{1}{\sqrt{r^2+|q|^2}}
       \left(\begin{array}{c}re^{i\theta}\\ q\end{array}\right)e^{\eta/2},
\ee
where we have set $z-z_1=re^{i\theta}$. It describes a vortex-like structure
centered at $z=z_1$, with size and orientation determined by the complex
parameter $q$. Using Eq. (\ref{25}), we can determine the asymptotic
behavior of $\eta$:
\be \label{40}
e^\eta=1-\frac{4|q|^2}{r^4}-\frac{8|q|^2(8+|q|^2)}{r^6}
+O\left(\frac{1}{r^8}\right).
\ee
This shows that if $q\neq0$, scalar fields approach the vacuum values like
$O(1/r^4)$, contrary to the $q=0$ case for which the limit are approached
exponentially. For $q\neq 0$ the magnetic field shows also a power fall-off at
large $r$, being given by
\be \label{41}
f_{12}=\frac{2|q|^2}{r^4}+\frac{4|q|^2(8+|q|^2)}{r^6}
+O\left(\frac{1}{r^8}\right).
\ee
On the other hand, the behaviors for smaller $r$ are found as
\bea \label{42}
e^\eta&=&c_0+\frac{c_0}{4}\left(\frac{4}{|q|^2}-c_0(c_0-1)\right)r^2
       +O(r^4),\nonumber\\
f_{12}&=&-\frac{1}{2}c_0(c_0-1)
       -\frac{1}{8}(1-2c_0)\left(\frac{4}{|q|^2}-c_0(c_0-1)\right)r^2+O(r^4),
\eea
where $c_0$ is a constant (which is equal to 1 if $q=0$). Note that the
magnetic field is
nonzero at the origin for $q\neq 0$. Hence its behaviors are different from
those of Ref. \cite{hkp} at both short and large distances. Another interesting
feature is that the solution approximates a $CP^1$ lump when $|q|\gg 1$. This
was noted in Ref. \cite{gibbons} for the case {\bf A}, and we can apply the
same argument here. Namely, for $|q|\gg 1$, we have
$\rho=4|q|^2/(r^2+|q|^2)^2\approx 0$ and $\eta\approx 0$ so that the
scalar fields lie effectively on the vacuum manifold $S^3$,
\be \label{43}
\left(\begin{array}{c}\chi\\ \psi\end{array}\right)
\simeq\frac{1}{\sqrt{r^2+|q|^2}}
       \left(\begin{array}{c}re^{i\theta}\\ q\end{array}\right).
\ee
This represents a $CP^1$ lump and we may say that the topological solitons of
this model interpolate between Chern-Simons vortices and $CP^1$ lumps. We have
verified that the angular momentum $J$, given by the expression (\ref{35}),
approaches zero as we let $q\rightarrow\infty$.

Some numerical results are shown in Fig. 2.
In Fig 2(a) we plot rotationally symmetric solutions for the coupling ratios
$p\equiv \frac{e_3}{e_1}=1$ and 2, while choosing
$|\chi(\infty)|^2=|\chi_0|^2\equiv v^2/2e_1$. As in the case {\bf A}, sizes
become smaller for larger $p$ and the magnetic field gets confined in a
narrow ring when $p$ is
large. The values of the magnetic field are zero at the origin, which is a
characteristic feature of Chern-Simons vortices \cite{hkp}. In Fig. 2(b) we
plot
the solutions of the form (\ref{39}) for $|q|=0.2, 1, 5,$ and 10. It is
clear that the solutions approximate $CP^1$ lumps when
$|q|${\raisebox{-1.5mm}{$\stackrel{{\displaystyle >}}{\sim}$}}5
just as in the case {\bf A}. The magnetic field
here shows interesting behaviors; for small $|q|$
it is ring-shaped (as in Fig. 2(a)) even if it does not vanish at the origin,
while it looks similar to that in Fig. 1(a) for large $|q|$.
In this way, we see clearly that our solution indeed interpolates
Chern-Simons vortices and $CP^1$ lumps.

Now we turn to nontopological soliton solutions allowed in symmetric phase.
While Eq. (\ref{19}) is still
valid, there is no restriction on $n_1$ and $n_2$ because the vacuum
manifold is trivial. Let the asymptotic behaviors of scalar fields
be such that
\bea\label{45}
|\chi|& \sim & \frac{1}{r^\alpha},\nonumber\\
|\phi|&\sim & \frac{1}{r^\beta},
\eea
where $\alpha,$ $\beta$ are positive constants. Then, by Eq. (\ref{19}),
$\alpha$ and $\beta$ are not independent but related by
\be\label{46}
e_3(n_1 + \alpha)=e_1(n_2 + \beta).
\ee
This relation is due to the fact that there is only one gauge field in
this case. The flux is given by
\be\label{47}
\Phi=\frac{2\pi(n_1+\alpha)}{e_1}=\frac{2\pi(n_2+\beta)}{e_3}.
\ee
 At present there is no rigorous existence proof of nontopological solitons
in Chern-Simons theory even with one scalar field, although their existence is
strongly supported by numerical analysis. The situation is similar for the
present case. Here we expect that
the number of zero modes be larger than $4n$, as we have seen in the model
with one scalar field \cite{hkp}.

We have done some analysis on nontopological solitons in the case of
$e_1=e_3=e$, assuming the form (\ref{39}).
Even if we do not know the exact solution, we can find approximate solutions
for $|q|\gg 1$ by making use of the solutions for the corresponding
non-relativistic self-dual model.
Formally, in the non-relativistic limit, Eq. (\ref{22}) reduces to
\be\label{48}
\nabla^2\ln|\chi|^2+|\chi|^2+|\psi|^2=4\pi\sum_{r=1}^{n}\delta({\bf x}
      -{\bf x}_r),
\ee
Other than the Liouville-type solutions obtained by setting $|\chi|=|\psi|$,
the present author found that Eq. (\ref{48}) admits a family of
exact solutions \cite{klm} of the form
\be\label{49}
\left(\begin{array}{c}\chi\\ \psi\end{array}\right)
  =\frac{\sqrt{12}(P\partial_z Q-Q\partial_z P)}{(|P|^2+|Q|^2)^{3/2}}
    \left(\begin{array}{c}P\\ Q\end{array}\right),
\ee
where $P$, $Q$ are polynomials of $z$ sharing no common zeros and ${\bf x}_r$'s
(see the right hand side of Eq. (\ref{48})) can be identified with zeros of
$P(P\partial_z Q-Q\partial_z P)$. For our purpose, we take $P(z)=
e^{\pi i/3}q^{-\frac{1}{3}}_0z$ and $Q(z)=e^{\pi i/3}q^{\frac{2}{3}}_0$ so that
\be\label{50}
\left(\begin{array}{c}\chi\\ \psi\end{array}\right)
  =\frac{\sqrt{12}|q|}{(r^2+|q|^2)^{3/2}}
    \left(\begin{array}{c}re^{i\theta}\\ q\end{array}\right).
\ee
Eq. (\ref{50}) will be a good approximate solution of the relativistic system
when $|q|\gg 1$.
We have found the asymptotic behaviors $|\chi|\sim r^{-2}$ and
$|\psi|\sim r^{-3}$, and so we identify $\alpha=2$ and $\beta=3$ for this
solution. The gauge field $a_i$ is given by Eq. (\ref{34}) and $g(r)$ has
limiting values $g(\infty)=-5/2$ and $g(0)=-3/2$. With
these and using Eq. (\ref{35}), we can calculate the angular momentum for
this solution,
\bea  \label{52}
J&=&\frac{4\pi\kappa}{e^2}-\frac{3\pi\kappa}{e^2|q|^2}\nonumber\\
 &\simeq&\frac{4\pi\kappa}{e^2},\hspace{5mm}|q|\gg 1.
\eea
Therefore, in symmetric phase, we find a non-zero angular momentum even if
$|q|\gg 1$, contrary to the case in broken phase. Fig. 3(a) shows the plot
of rotationally symmetric nontopological soliton solutions
for $\alpha=2.1$ and $3$ when $p=2$. In Fig. 3(b) we plot solutions of the form
(\ref{39}), for $|q|=5$ and 10.
Note that for $|q|=10$, it is well approximated by the exact solutions of
nonrelativistic self-duality equations. It gives $\alpha=2.056$,
which is not far from the nonrelativistic value $\alpha=2$.

Closing this subsection we remark two things. First, this model may be viewed
as a linearized version of the $CP^1$ model with the Chern-Simons term.
The potential has been chosen by demanding self-duality, and then the
symmetric phase becomes degenerate with the asymmetric
phase. In this sense, this model interpolates
two popular field-theoretic models useful for high-$T_c$
superconductivity \cite{polyakov}, namely, the $CP^1$ model and the
Chern-Simons-Higgs model. Second, while the case {\bf A} interpolates
Ginzburg-Landau
vortices and $CP^1$ lumps, the case we just discussed interpolates
Chern-Simons vortices and $CP^1$ lumps (in broken phase).
Then it is natural to expect that $U(1)$ Maxwell-Chern-Simons
theory \cite{llm} with two complex scalar fields interpolates all three
kinds of solitons. This model is briefly discussed in the appendix.

\subsection*{C. {\boldmath $e_3=e_4=0$}}
\ \indent
In this case, the field $\psi$ decouples from the rest and
the potential reduces to
\be  \label{53}
U=\frac{1}{2}(e_2|\chi|^2-u^2)^2
	  +\frac{e_1^2}{\kappa^2}|\chi|^2
  \left(e_1|\chi|^2-\frac{e_1}{e_2}u^2-\frac{\kappa e_2}{e_1}N\right)^2,
\ee
where we have adjusted the field $N$ by a suitable constant shift.
Assuming $e_2>0$, the vacuum configuration then clearly corresponds to
\be  \label{54}
|\chi|^2=u^2/e_2,\hspace{5mm} N=0.
\ee
So only the broken phase is possible. This case is different from the previous
two cases in that both Maxwell and Chern-Simons gauge fields couple to a
single complex scalar $\chi$ simultaneously. From Eqs. (\ref{5}) and (\ref{6}),
we note that
\be  \label{55}
\kappa\Phi_a=e_1 Q_\chi=0.
\ee
Therefore, there is no charged vortex as noted in Ref. \cite{llm}. If $e_2=0$,
this model reduces to the pure Chern-Simons model which is known to have
charged vortices. Hence, $e_2=0$ is a singular point, i.e., the theory
with $e_2=0$ has distinct behaviors from that with $e_2\neq 0$.

We find it convenient to work with dimensionless quantities by making
the replacements
$$
x_\mu\rightarrow\frac{x_\mu}{\sqrt{2e_2}u},\hspace{5mm}
\chi\rightarrow\frac{u}{\sqrt{e_2}}\chi,\hspace{5mm}
N\rightarrow\frac{e_1^2u^2}{\kappa e_2^2}N,
$$
$$
a_\mu\rightarrow\frac{\sqrt{2e_2 u}}{e_1}a_\mu,\hspace{5mm}
A_\mu\rightarrow\sqrt{\frac{2}{e_2}}uA_\mu.
$$
Then, after eliminating gauge fields, self-duality equations yield the
following equation for $|\chi|$ and $N$ (here $\xi=e_1^4u^2/\kappa^2e_2^3$):
\bea  \label{56}
\nabla^2\ln|\chi|^2-2\xi|\chi|^2(|\chi|^2-N-1)-|\chi|^2+1
       &=&4\pi\sum_{r=1}^{n}\delta({\bf x}-{\bf x}_r),\nonumber\\
\nabla^2N-|\chi|^2(|\chi|^2-N-1)&=&0,
\eea
For these coupled equations, no analytic solution is available at present.
For rotationally symmetric solutions, we write
\bea  \label{57}
|\chi|=f(r)e^{in\theta}, &&N=N(r), \nonumber\\
a_i=\frac{\varepsilon^{ij}x^j}{r^2}g_1(r),&&
A_i=\frac{\varepsilon^{ij}x^j}{r^2}[g_2(r)-n],
\eea
with various radial functions here satisfying the boundary conditions
\bea  \label{58}
nf(0)&=&g_1(0)=N'(0)=0,\nonumber\\
g_2(0)&=&n,\nonumber\\
g_1(\infty)&=&g_2(\infty)=h(\infty)=0,\\
f(\infty)&=&1.\nonumber
\eea
For these solutions, the angular momentum is given as
\bea  \label{59}
J&=&-\frac{1}{\sqrt{2e_2^3}}\inttd \varepsilon^{ij}x^i(D_0\chi^*D_j\chi
   +D_0\chi D_j\chi^*+\varepsilon^{jk}F_{0k}F_{12})\nonumber\\
 &=&-\frac{1}{\sqrt{2e_2^3}}\inttd \varepsilon^{ij}x^i
   \left[\frac{\varepsilon^{jk}x^k}{r^3}g_1'(g_1+g_2)
	 +\frac{\varepsilon^{jk}x^k}{r^3}g_1g_2'\right]\nonumber\\
 &=&\frac{1}{\sqrt{2e_2^3}}\intr 2\pi \left(\frac{1}{2}g_1^2+g_1g_2\right)'
                \nonumber\\
 &=&0.
\eea
This null result is expected since there is no charged vortices in this model.

In Fig. 4, some numerical results are given for rotationally symmetric
solutions with $n=1$. Note that the profiles of $|\chi|$ and $F_{12}$ are
similar to those found in the ordinary abelian Higgs model. The
Chern-Simons magnetic field $f_{12}$ shows an interesting behavior. It is
positive for small $r$, then changes sign
and goes to 0 as $r\rightarrow\infty$. This is expected because there is no
charged vortex, i.e., $\Phi_a=0$, as we noted above.

The following comment on this model may be useful. One may view this model as
the ordinary self-dual abelian Higgs model modified by the addition of
the Chern-Simons term, and the effect is to make the neutral scalar $N$
nontrivial. But the main features of this model are essentially the same as
those of the abelian-Higgs model: there is only broken phase, and
no charged or spinning vortex exists.

\subsection*{D. {\boldmath $e_1=e_3, e_2=e_4$}}
\ \indent
Here, two complex scalar fields $\chi$ and $\psi$ enter the theory in such a
way that an additional global $SU(2)$ symmetry may be present. The potential is
\be  \label{60}
U=\frac{1}{2}(e_2|\Psi|^2-u^2)+\frac{e_1}{\kappa^2}|\Psi|^2\left(e_1|\Psi|^2-
  \kappa\frac{e_2}{e_1}N\right),
\ee
where $\Psi=\left(\begin{array}{c}\chi\\ \psi\end{array}\right)$.
The solution of this case will possess properties corresponding to a
juxtaposition of the cases {\bf A} and {\bf C}; in particular,
there is no charged or spinning vortex solution. As we have seen in the cases
{\bf A} and {\bf B}, $SU(2)$ global symmetry enables us to write
$ w(z)=\frac{\psi}{\chi}=\frac{Q_n(z)}{P_n(z)}$, where $P_n,$ $Q_n$ are the
same as before. Then the resulting self-duality equations can be read from
Eqs. (\ref{25}) and (\ref{56}),
\bea  \label{61}
\nabla^2\eta-2\xi e^\eta(e^\eta-N-1)-e^\eta+1&=&\rho,\nonumber\\
\nabla^2N-e^\eta(e^\eta-N-1)&=&0,
\eea
where $\eta$ and $\rho$ are defined by Eqs. (\ref{24}) and (\ref{26}). In this
case, however, a rigorous existence proof for $\rho=0$ case
is not available and, strictly speaking, it is just a conjecture
that the solution exists for each $P_n(z)$ and $Q_n(z)$ in Eq. (\ref{31}).

Without going further on this model, we just make one comment. As before
we may consider this model as a linearized version of $CP^1$ model to which a
Chern-Simons gauge field is coupled. A neutral field $N$ appears to make the
system self-dual. From the self-duality equations (\ref{61}),
it is easy to see that for solutions of the form Eq. (\ref{39}),
\bea  \label{62}
u&\simeq & 0,\nonumber\\
N&\simeq & 0 \hspace{7mm}\mbox{when }|q|\gg 1
\eea
and they again approximate $CP^1$ lumps as expected.

\subsection*{E. {\boldmath $e_1=e_3,$ $e_2=-e_4$} and {\boldmath $u^2=0$}}
\ \indent
Finally, we will consider the case with $e_1=e_3$ and $e_2=-e_4$, while setting
$u^2=0$. This case is interesting because both gauge fields couple to
scalar fields in a nontrivial manner, and there are both
symmetric and asymmetric phases which are degenerate. The potential reads
\bea  \label{63}
U&=&\frac{e_2^2}{2}(|\chi|^2-|\psi|^2)^2+\frac{e_1^2}{\kappa^2}|\chi|^2
  \left[e_1(|\chi|^2+|\psi|^2)-\frac{\kappa e_2}{e_1}N-v^2\right]^2\nonumber\\
&& +\frac{e_1^2}{\kappa^2}|\psi|^2
  \left[e_1(|\chi|^2+|\psi|^2)+\frac{\kappa e_2}{e_1}N-v^2\right]^2
\eea
and we have a symmetric phase if the vacuum corresponds to
$|\chi|=|\psi|=0$ and $N=$(an arbitrary constant), and an asymmetric phase
with the vacuum described by $|\chi|^2=|\psi|^2=v^2/2e_1$ and  $N\equiv 0$.

To make the fields dimensionless, we make the replacements
$$
x_\mu\rightarrow\frac{\sqrt{e_1}}{e_2v}x_\mu,\hspace{5mm}
\chi\rightarrow\frac{v}{\sqrt{2e_1}}\chi,\hspace{5mm}
\psi\rightarrow\frac{v}{\sqrt{2e_1}}\psi,\hspace{5mm}
$$
$$
N\rightarrow\frac{e_1^2v^2}{2\kappa e_2}N,\hspace{5mm}
a_\mu\rightarrow\frac{e_2 v}{e_1^{3/2}}a_\mu,\hspace{5mm}
A_\mu\rightarrow\frac{v}{\sqrt{e_1}}uA_\mu.
$$
Then after eliminating gauge fields from the  self-duality equations we obtain
the following set of equations:
\bea  \label{64}
\!\!\!\nabla^2\ln|\chi|^2\!\!\!&=&\!\!\!\xi[(\chi|^2+|\psi|^2)(\chi|^2
 +|\psi|^2-2)(|\chi|^2-|\psi|^2)N]+|\chi|^2-|\psi|^2\nonumber\\
       &&+4\pi\sum_{r=1}^{n_1}\delta({\bf x}-{\bf x}_r),\nonumber\\
\!\!\!\nabla^2\ln|\psi|^2\!\!\!&=&\!\!\!\xi[(\chi|^2+|\psi|^2)(\chi|^2
+|\psi|^2-2)(|\chi|^2-|\psi|^2)N]+|\chi|^2-|\psi|^2\\
       &&+4\pi\sum_{r=1}^{n_2}\delta({\bf x}-{\bf x'}_r),\nonumber\\
\!\!\!\nabla^2N\!\!\!&=&\!\!\!-(|\chi|^2-|\psi|^2)(\chi|^2+|\psi|^2-2)+
N(|\chi|^2+|\psi|^2) \nonumber,
\eea
where $\xi\equiv e_1^3v^2/\kappa^2e_2^2$. Being nontrivially coupled and
highly nonlinear, analytic solutions to these equations appear to be out of
question. But with the ansatz
\be  \label{65}
|\chi|=|\psi|,\hspace{5mm}N=0,
\ee
we can reduce the above three equations to the following single equation
$$
\nabla^2\ln|\chi|^2-4\xi|\chi|^2(|\chi|^2-1)
       =4\pi\sum_{r=1}^{n_1}\delta({\bf x}-{\bf x}_r),
$$
which is identical to that appearing in the self-dual Chern-Simons model
with one scalar field \cite{hkp}. (See also Eq. (\ref{22}).) It is unclear
whether there exist
solutions other than these. Especially, it will be interesting to see
whether there is a solution with half integer vorticity for both $\Phi_a$ and
$\Phi_A$ \cite{llm}.

Before investigating some of these issues, we first calculate quantum
numbers of rotationally symmetric configurations, which are specified as
\bea  \label{66}
\chi&=&f_1(r)e^{in_1\theta},\hspace{30mm}\psi=f_2(r)e^{in_2\theta},\nonumber\\
a_i&=&\varepsilon^{ij}\frac{x^j}{r^2}\left(g_1(r)-\frac{n_1+n_2}{2}\right),
\hspace{3mm}
A_i=\varepsilon^{ij}\frac{x^j}{r^2}\left(g_2(r)-\frac{n_1-n_2}{2}\right),\\
N&=&N(r).\nonumber
\eea
To ensure finite energy, various functions here are subject to following
boundary conditions: at spatial infinity,
\bea  \label{67}
f_1(\infty)&=&f_2(\infty)=1 \,\,\mbox{ or }0,\nonumber\\
g_1(\infty)&=&-\frac{\alpha+\beta}{2},
        \hspace{3mm}g_2(\infty)=-\frac{\alpha-\beta}{2},\\
N(\infty)&=&N_0,\nonumber
\eea
while, at $r=0$,
\bea  \label{68}
n_1f_1(0)&=&n_2f_2(0)=N'(0)=0,\\
g_1(0)&=&\frac{n_1+n_2}{2},
        \hspace{3mm}g_2(0)=\frac{n_1-n_2}{2}.\nonumber\\
\eea
[In Eq. (\ref{67}), we must set $\alpha=\beta=N_0=0$ for solutions in the
broken phase.] Then, as usual, we find the flux values
\bea  \label{69}
\Phi_a=2\pi\left(\frac{n_1+n_2}{2}+\frac{\alpha+\beta}{2}\right),\nonumber\\
\Phi_A=2\pi\left(\frac{n_1-n_2}{2}+\frac{\alpha-\beta}{2}\right),
\eea
while the angular momentum $J$ is given by
\bea  \label{70}
J&=&-\frac{v}{2\sqrt{e_1}e_2}\inttd \varepsilon^{ij}x^i(D_0\chi*D_j\chi
   +D_0\psi*D_j\psi+c.c.+2\varepsilon^{jk}F_{0k}F_{12})\nonumber\\
 &=&\frac{v}{2\sqrt{e_1}e_2}\inttd \left(\frac{2}{\sqrt\xi}g_1\frac{g'_1}{r}
   +2g_2\dlz F^{i0}-2F_{i0}\frac{x^i}{r}g'_2\right)\nonumber\\
 &=&\frac{2\pi v}{\sqrt{e_1}e_2}\left.\left(\frac{g_1^2}{2\sqrt\xi}
   - g_2 r A'_0\right)\right|^\infty_0\nonumber\\
 &=&-\frac{Q^2}{\pi\kappa}+\frac{\alpha+\beta}{e_1}Q\nonumber\\
 &=&\frac{Q^2}{\pi\kappa}-\frac{n_1+n_2}{e_1}Q,
\eea
where $Q=Q_\chi+Q_\psi$(=$2Q_\chi$, from Eq. (\ref{5})) denote the total
electric charge.
Note that $J$ does not depend on couplings associated with the the Maxwell
field; this is natural since Chern-Simons interaction is responsible for
non-zero angular momentum.

Finally we turn to self-duality equations in Eq. (\ref{64}) and discuss the
possibility of vortex solutions with half-integer vorticity (i.e.
$\Phi_a =2\pi(\frac{1}{2}+\frac{\alpha +\beta}{2})$ and
$Phi_A=2\pi(\frac{1}{2}+\frac{\alpha-\beta}{2})$).
Since Eq. (\ref{64}) is very complicated we made
some numerical studies assuming the rotationally symmetric ansatz (\ref{66})
with $n_1=1$, $n_2=0$. Specifically we
tried relaxation methods using the package COLSYS \cite{colsys} as well as
shooting methods, both unidirectional and bidirectional \cite{recipie}.
But we failed to find such solutions for both topological and nontopological
cases. A plausible reason for our negative finding might be that the
boundary condition $N(\infty)$ may well be inconsistent with one of the
Gauss laws which requires that $Q_\chi=Q_\psi$. As we shoot the
fields near the origin, we found that $N$ is either logarithmically divergent
or
becomes singular at a finite value of $r$. [Note that with the ansatz
(\ref{65}), $Q_\chi=Q_\psi$ is automatically satisfied.]
Of course, as we have already seen in one instance (at the end of the
subsection {\bf A}), that we have appropriate self-duality equations does not
imply the actual existence of a non-trivial solution for which the given energy
bound is saturated. In spite of what we have observed, however, there
remains some possibility
that such solutions do exist and our numerical analysis is incomplete; for
example, solutions may be found with initial parameters which we did not
sweep. Therefore, it is a bit hasty to conclude that there is no vortex
solution with half-integer vorticities in this model.

\section{SUMMARY}
\ \indent
In this paper we have studied various interesting special cases in the
$U(1)\times U(1)$ self-dual Maxwell-Chern-Simons model. They
include abelian Higgs system with two complex scalar fields
(the case {\bf A}) and its Chern-Simons counterpart (the case {\bf B}).
The latter can be regarded as a linearized version of the $CP^1$ model with
the Chern-Simons term (while maintaining self-duality) and as such
it interpolates two familiar field theory systems appearing
in the discussion of high-$T_c$ superconductivity \cite{polyakov}.
We proved the existence
of multisoliton solutions for the topological soliton case, along the lines
taken in Refs. \cite{gibbons} and  \cite{wang}.
In the symmetric phase we discussed the properties
of nontopological soliton solutions by utilizing some known exact solutions
of the corresponding nonrelativistic model.
A more general field theory model interpolating the self-dual systems
of the cases {\bf A} and {\bf B} is given in the appendix.

With $e_3=e_4=0$, the resulting self-dual model consists of a complex
scalar, a neutral scalar, the Maxwell field, and the Chern-Simons field.
It has no charged vortex because of the Gauss law and has more or less
the same properties as those of the ordinary abelian Higgs model.
The role of the Chern-Simons field is to
modify field configurations such that neutral field $N$ may have
nontrivial values.
Extension of this case to a system with two complex scalar fields was also
discussed.

Finally we considered a model where all scalar fields and gauge fields couple
to one another nontrivially. We calculated the quantum numbers for rotationally
symmetric configurations, and showed that the ansatz $|\chi|=|\psi|$ leads back
to the self-duality equations found in the `minimal' self-dual Chern-Simons
system \cite{hkp}. We performed numerical analysis to find vortex solutions
with half-integer vorticities and the result was negative. However,
there is still a room for further examination.

As we mentioned in the introduction, this generalized abelian Higgs model
with independent Chern-Simons gauge term may have some relevance in describing
real physical systems. Then a variety of vortex structures we discussed in
this paper are expected to have some useful physical applications. These issues
are under investigation.

\section*{ACKNOWLEDGMENTS}
\ \indent
It is a great pleasure to thank Prof. Choonkyu Lee who suggested this
problem and carefully read the manuscript. Useful conversations with Prof.
Kimyeong Lee are also acknowledged. This work was supported in part by the
Korea Science and Engineering Foundation (through the Center for Theoretical
Physics, SNU) and by the Ministry of Education, Korea.

\section*{APPENDIX}
\ \indent
Here, we briefly describe the self-dual topologically massive $U(1)$ gauge
theory with two complex scalar fields and a neutral one, extending the
model of Ref. \cite{llm} such that the model may have an additional global
$SU(2)$ symmetry. The Lagrangian is given by
\be
\lagrangian = -\frac{1}{4} \flmn^2 +\frac{1}{4}\eumnr\flmn A_\rho +
 |D_\mu\Psi|^2 + \frac{1}{2}(\dlm N)^2-U(\Psi,N),
 \ee
where
$$
D_\mu\Psi=(\dlm-ie A_\mu)\left(\begin{array}{c}\phi_1\\
           \phi_2\end{array}\right),
$$
and
\be
U(\Psi, N)=\frac{1}{2}(e|\Psi|^2+\kappa N-ev^2)^2+e^2N^2|\Psi|^2
\ee
with
$$
|\Psi|^2=|\phi_1|^2+|\phi_2|^2.
$$
It has global $SU(2)$ symmetry in addition to local $U(1)$ symmetry.
Gauss law reads
\be
-\dli F_{i0}+\kappa F_{12}-eJ^0=0,
\ee
where
$$
J_0=-i(\Psi^\dagger D_0\Psi-D_0\Psi^\dagger\Psi).
$$
Integrating over space, we obtain the usual relation
\be
\kappa\Phi=eQ,
\ee
where $Q=\inttd J_0$. It is not difficult to show that the energy functional
\be
E=\inttd \left[\frac{1}{2}F_{i0}+\frac{1}{2}F_{12}^2+|D_0\Psi|^2+|D_i\Psi|^2
+\frac{1}{2}(\dlz N)^2+\frac{1}{2}(\dli N)^2+U\right]
\ee
satisfies the following inequality
\be
E\ge|ev^2\Phi|,
\ee
where the equality holds if and only if the following self-duality
equations are satisfied (static fields assumed):
\bea
(D_1\pm i D_2)\Psi&=&0,\nonumber\\
F_{i0}\pm\dli N&=&0,\nonumber\\
(D_0\mp ieN)\Psi&=&0,\\
F_{12}\pm(\kappa N+e|\Psi|^2-ev^2)&=&0.\nonumber
\eea
For a finite energy, we must here demand that
$$
r\rightarrow \infty:\hspace{5mm}
|\Psi|^2\lrarrow v^2,\hspace{5mm} N\lrarrow 0\hspace{7mm}
          \mbox{asymmetric phase,}
$$
or
$$
r\rightarrow \infty:\hspace{5mm}
|\Psi|\lrarrow 0, \hspace{5mm} N\lrarrow\frac{ev^2}{\kappa}\hspace{7mm}
          \mbox{symmetric phase.}
$$
In the asymmetric phase, the vacuum manifold is $S^3$.
Therefore, as in the case {\bf A} of Sec. III, the phase of the scalar
field $\Psi$ fibers the vacuum manifold as a $U(1)$ bundle
over $CP^1\simeq S^2$, and so a given sector of the theory is specified by
a choice of a point on $S^2$ and the winding number of the scalar field around
the fiber over that point.

If only one complex scalar field were present, the model would interpolate the
Ginzburg-Landau model and the pure Chern-Simons model (as noted in Ref.
\cite{llm}). Hence, in the case of the above model (with $SU(2)$ global
symmetry), it should be clear that the case {\bf A} and the case {\bf B} of
Sec. III can be obtained in the appropriate limiting cases. In other words,
the solutions in this model will interpolates three kinds of
solitons solutions, i.e., Ginzburg-Landau vortices, Chern-Simons vortices
and $CP^1$ lumps.

\section*{Figure captions}
\begin{description}
\item[{\rm Figure 1.}] Plot of solutions for the case {\bf A}.
(a) Rotationally symmetric solutions with $n=1$, for $p(\equiv e_4/e_2)=1$
(dotted lines) and for $p=2$ (solid line).
We choose $|\chi(\infty)|^2=|\chi_0|^2\equiv u^2/2e_2$.
Lines approaching zero asymptotically are magnetic fields $F_{12}$.
The solid line approaching 1
is $|\chi|$, that approaching $1/\sqrt{2}$ is $|\psi|$, and dashed line
approaching 1 is $|\chi|(=|\psi|)$ for $p=1$.
(b) The functions $e^\eta/2$ (approaching 1) and $F_{12}$ (approaching 0)
for $n_1=1$ and $n_2=0$ in the $e_2=e_4=e$ case. The dotted, solid,
dashed, and dot-dashed lines correspond to $|q|=0.2,$ 1, 5, and 10,
respectively.

\vspace{.5cm}
\item[{\rm Figure 2.}] Plot of solutions in an asymmetric phase for the
case {\bf B}.
(a) Rotationally symmetric solutions with $n=1$ for $p(\equiv e_3/e_1)=1$
(dotted lines) and for $p=2$ (solid line). We choose
$|\chi(\infty)|^2=|\chi_0|^2\equiv v^2/2e_1$.
Lines approaching zero asymptotically are magnetic fields $f_{12}$.
The solid line approaching 1
is $|\chi|$, that approaching $1/\sqrt{2}$ is $|\psi|$, and the dashed line
approaching 1 is $|\chi|(=|\psi|)$ for $p=1$.
(b) $e^{\eta/2}$ (approaching 1) and $f_{12}$ (approaching 0) for
solutions having the form in Eq. (\ref{39}).
Dotted, solid, dashed, and dot-dashed lines correspond to $q=0.2$, 1, 5, and
10, respectively.

\vspace{.5cm}
\item[{\rm Figure 3.}] Plot of solutions in the symmetric phase for the case
{\bf B}. (a) Solid lines are for $n=0$ and $\alpha=2.1$ with
$p\equiv e_3/e_1=2$. Dashed lines are for $n=1$ and
$\alpha=3$ with $p=2$. Among the same types of lines having values less
than 1, upper lines
correspond to $|\chi|$ and the other to $|\psi|$. The magnetic field
$f_{12}$ is ring-shaped in both cases.
(b) The functions $e^{\eta/2}$ (starting at $\frac{\sqrt{12}}{10}$)
and $f_{12}$ for solutions of the form in Eq. (\ref{39}).
The dashed line is for $|q|=5$, and solid line for $|q|=10$.
The dotted line is the plot of the function $\sqrt{12}|q|/(r^2+|q|^2)$
for $|q|=10$, a solution of nonrelativistic self-duality equations.

\vspace{.5cm}
\item[{\rm Figure 4.}] Plot of rotationally symmetric solutions with unit
vorticity for the case {\bf C}.
Dashed and solid lines are for $\xi=1$ and 2, respectively. $|\chi|$ are
represented by lines approaching 1, and $N$ by negative definite lines. Lines
starting at 1 represent $F_{12}$; lines having both positive and negative
values represent $f_{12}$.
\end{description}

\newpage

\begin{thebibliography}{99}
\bibitem{polyakov}A. Fetter, C. Hanna, and R. Loughlin, Phys. Rev. B {\bf 39},
9679(1989); Y. Chen {\it et al.} Int. J. Mod. Phys. B {\bf 3}, 1001(1989);
X. Wen and A. Zee, Nucl. Phys. B (Proc. Suppl.), {\bf 15}, 135 (1990);
T. Banks and J. Lykken, Nucl. Phys. B {\bf 336}, 500 (1990);
J. Lykken, J. Sonnenschein, and N. Wess, Int. J. Mod. Phys. A {\bf 6},
5155 (1991);
I. E. Dzyaloshinskii, A. M. Polyakov, and P. B. Wiegmann, Phys Lett A {\bf 127}
(1988) 112.
\bibitem{khare}S. Paul and A. Khare, Phys. Lett B {\bf 174}, 420 (1986);
{\bf 182}, 414(E) (1986);
S. Paul and A. Khare, Phys Lett. B {\bf 236}, 283 (1989).
\bibitem{cp1}P. Voruganti, Phys. Lett. B {\bf 223}, 181 (1989);
M. A. Mehta, J. A. Davies, and I. J. R. Aitchson,
Phys. Lett. B {\bf 281}, 86, (1992).
\bibitem{hkp}J. Hong, Y. Kim, and P.-Y. Pac,
Phys. Rev. Lett. {\bf 64}, 2230 (1990);
R. Jackiw and E. Weinberg, Phys. Rev. Lett {\bf 64}, 2234 (1990);
R. Jackiw, K. Lee, and E. Weinberg, Phys. Rev. D {\bf 42}, 3488 (1990).
\bibitem{jp}R. Jackiw and S.-Y. Pi, Phys. Rev. Lett. {\bf 64}, 2969 (1990);
 Phys. Rev. D {\bf 42}, 3500 (1990); K. Lee, Phys. Rev. Lett. {\bf 66},
553 (1991); K. Lee, Phys. Lett. B {\bf 255}, 381 (1991).
\bibitem{llm}C. Lee, K. Lee, and H. Min, Phys. Lett. B {\bf 252}, 79 (1990).
\bibitem{lmr}C. Lee, H. Min, and C. Rim, Phys. Rev. D {\bf 43}, 4100 (1991).
\bibitem{gibbons}T. Vachaspati and A. Ach\`{u}carro, Phys. Rev. D {bf 44}
3067 (1991); M. Hindmarsh, Phys. Rev. Lett. {\bf 68}, 1263  (1992);
G. W. Gibbons, M. E. Ortiz, F. Ruiz Ruiz, and T. M. Samols, CTP\# 2063.
\bibitem{wang}R. Wang, Commun. Math. Phys. {\bf 137}, 587 (1991).
\bibitem{coleman}See for example S. Coleman, {\it Aspects of Symmetry},
Cambridge University Press (1985).
\bibitem{taubs}C. H. Taubes, Commun. Math. Phys. {\bf 72} 277 (1980);
A. Jaffe and C. H. Taubes, {\it Vortices and Monopoles}, Birkhauser, Boston
(1980).
\bibitem{klm}C. Kim, C. Lee, and H. Min, in preparation.
\bibitem{landau}A. A. Abrikosov, Sov. Phys. JETP {\bf 5}, 1174 (1957);
H. B. Nielsen and P. Olesen, Nucl. Phys. B {\bf 61}, 45 (1973).
\bibitem{weinberg}E. J. Weinberg, Phys Rev. D {\bf 19}, 3008 (1979).
\bibitem{colsys}U. Ascher, J. Christiansen, and R. D. Russell, ACM Trans. Math.
Softw. {\bf 7}, 209 (1981).
\bibitem{recipie}W. H. Press, B. P. Flannery, S. A. Teukolsky, and W. T.
Vetterling, {\it Numerical Recipes}, Cambridge University Press (1986)
\end{thebibliography}
\end{document}